\documentclass[twoside,reqno]{bjp}
\usepackage{epsfig,cite}
\usepackage{latexsym}
\usepackage{graphicx}
\usepackage{color}
\usepackage{amssymb,amsmath}
\usepackage{times}
\usepackage{setspace}
\begin{document}

\sloppy \raggedbottom

\setcounter{page}{1}



\title{Frenkel Excitons--Charge-Transfer Excitons--Phonons Coupling in One-Component Molecular Crystals\thanks{This work is dedicated to the memory of Academician Matey Mateev (1940--2010), a great friend and wonderful colleague on Bulgarian physics community.}}

\runningheads{Excitons--CTEs--Phonons Coupling in \ldots}{I.~J.~Lalov,
I.~Zhelyazkov}

\begin{start}
\author{I.~J.~Lalov}{}, \coauthor{I.~Zhelyazkov\thanks{Corresponding author: \texttt{izh@phys.uni-sofia.bg}}}{}

\address{Faculty of Physics, Sofia University, BG-1164 Sofia, Bulgaria}{}

\received{21 February 2011}

\begin{Abstract}
In this paper, we simulate the linear absorption spectra of the MePTCDI and PTCDA crystals.  The basic Hamiltonian describes the Frenkel excitons and charge-transfer excitons mixing in the molecular stack (point group $C_i$) and their linear coupling with one vibrational mode of an intramolecular vibration.  Using the vibronic approach, we calculate the linear optical susceptibility in the excitonic and one-phonon vibronic regions of the molecular stack and of a crystal with two types of nonequivalent stacks.  We put the excitonic and vibrational parameters for the two crystals fitted in the previous studies and analyze some important features of of the linear absorption lineshape in the spectral region of $15\,000$--$23\,000$ cm$^{-1}$ and the virtual positions of many-particle  bands.  Our study exhibits the necessity of introducing the FE--CTEs mixing in the interpretation of the linear absorption spectra, especially in the MePTCDI crystal.
\end{Abstract}

\end{start}

\section{Introduction}
\label{sec:intro}
The mixing of Frenkel excitons (FEs) and charge-transfer excitons (CTEs) has been studied both experimentally and theoretically in many one-compo\-nent molecular crystals, \emph{e.g.}, polyacenes \cite{sebastian81,sebastian83,siebrand83,petelenz96}, perylene derivatives \cite{henessy99,hoffmann00,hoffmann02}, fullerenes \cite{jeglinski92,pac98}, and other.  The excitonic and vibronic spectra of quasi-one-dimensional crystals like 3,4,9,10-perylenetetracarboxylic dianhydride (PTCDA) and $N,N^{\prime}$-dimethylperylene-3,4,9,10-perylenetetracarbo\-ximide (MePTCDI) have been treated in \cite{henessy99,hoffmann00,hoffmann02} and \cite{schmidt02,lalov06a,lalov07a,lalov06b,lalov06c}.  The short molecular distance in quasi-one-dimensional stacks causes a strong FE--CTEs mixing and mixing of their vibronic satellites as well, especially in the absorption spectra in the spectral region of $2$--$3$ eV ($15\,000$--$23\,000$ cm$^{-1}$).

In the present paper, we calculate the linear absorption spectra of one-component MePTCDI and PTCDA crystals applying the vibronic approach  developed in our previous papers \cite{lalov06b,lalov07b}.  It allows complex calculations of the absorption in pure excitonic, one-phonon vibronic, two-phonon vibronic, etc.\ spectra.  In paper \cite{lalov08} the vibronic approach is the tool of studying the FE--CTEs mixing in a two-component stack of alternatively arranged donor--acceptor (DA)-molecules.  While the FE--CTEs mixing in \cite{lalov08} is a probable but hypothetical model, in the present study we turn to the real absorption spectra and use the parameters of the FE--CTEs mixing from Refs.~\cite{henessy99,hoffmann00,schmidt02}.

Two general differences exhibit the models of the FE--CTEs mixing in one- and two-component molecular stacks \cite{lalov08}, notably: (i) The intermolecular transfer both of electrons and holes must be a feature of the model for the case of one-component stack whereas only one transfer mechanism is sufficient in a DA-molecular stack.  In the last case the Frenkel exciton represents a collectivized electronic excitation of a donor or an acceptor and the strongest transfer on the closest neighbor would be either of a hole or correspondingly of an electron.  Obviously in an one-component stack both types of transfer on the neighbors could be of the comparable probability.  The two-step processes of successive transfer of the electron and the hole on the same molecule ensure the transfer of a FE even in the case of its relatively weak direct intermolecular transfer \cite{schmidt02}.  In this way, our present study extends the calculations in \cite{lalov06b} where only one transfer mechanism has been considered. (ii) In the most widely studied DA-crystal, anthracene-PMDA the excitonic absorption lines are narrow \cite{haarer75,brillante80,weiser04} and many details in the vibronic spectra can be seen.  In PTCDA and MePTCDI crystals the absorption lines are two order of magnitude wider.  Thus we pay attention to the general structure of the excitonic and one-phonon vibronic spectra supposing a width of the excitonic lines of $300$--$500$ cm$^{-1}$ (not $2$--$10$ cm$^{-1}$ which is the absorption width of the anthracene-PMDA).

In the next section of the paper we involve the initial Hamiltonian in the case of a FE--CTEs mixing.  The Hamiltonian contains one mode of intramolecular vibration linearly coupled to the FE and CTEs.  In Section~3 the linear optical susceptibility has been calculated in the excitonic and one-phonon vibronic spectra.  In Section~4 the linear absorption has been modelled using the excitonic and vibrational parameters of the PTCDA and MePTCDI crystals fitted in Refs.~\cite{hoffmann00,hoffmann02,schmidt02}.  Section~5 summarizes the main findings derived in this paper.  In the Appendix we treat some problems of the FE--CTEs mixing in a molecular stack of the $C_i$ point group of symmetry.

\section{Hamiltonian for the Case of a FE--CTEs--Phonon Coupling}
\label{sec:Hamiltonian}
We consider the excitonic and vibronic excitations in a linear molecular stack of $N$ identical molecules which are regularly arranged at a distance $d$ each other.  The point group of symmetry of the stack is $C_i$ as is for the molecular stacks of PTCDA and MePTCDI \cite{hoffmann00}.  The origin of the Frenkel exciton is a non-degenerate molecular electronic excitation with excitation energy $E_{\rm F}$ and $L$ being the transfer integral between neighboring molecules.  We denote by $B_n$ ($B_n^{+}$) the annihilation (creation) operator of the electronic excitation on molecule $n$ and get the following FE-part of the Hamiltonian:
\begin{equation}
\label{eq:fehamilt}
    H_{\rm FE} = \sum_n E_{\rm F} B_n^{+}B_n + \sum_{n,n^{\prime}} L \left( \delta_{n^{\prime},n+1} + \delta_{n^{\prime},n-1} \right) B_{n^{\prime}}^{+}B_n.
\end{equation}
As usually, we consider two CTEs of equal excitation energy $E_{\rm c}$ and $C_{n,1}$ is the annihilation operator CTE, $\sigma =1$, with hole located on the site $n$ and electron on the site $n + 1$, whereas the electron of the second CTE, $\sigma =2$, ($C_{n,2}$) is located at molecule $n - 1$.  We neglect the transfer of CTEs as a whole and the mutual coupling of the two CTEs since those processes can be realized through the transfer of the electron or hole at distance $2d$ which is less probable than the FE--CTEs mixing caused by the electron (hole) transfer at the neighbor molecule (see Refs.~\cite{lalov08,brillante80}).  We have the following CTEs-part of the Hamiltonian:
\begin{equation}
\label{eq:cteshamilt}
    \hat{H}_{\rm CTE} = \sum_{n,\sigma = 1,2} E_{\rm c}C_{n \sigma}^{+}C_{n \sigma}
\end{equation}
and suppose the following operator for the FE--CTEs mixing:
\begin{eqnarray}
\label{eq:fcoperator}
    \hat{H}_{\rm FCT} = \sum_n \left[ \varepsilon_{\rm e1}B_n^{+}C_{n,1} + \varepsilon_{\rm e2}B_n^{+}C_{n,2} \right. \nonumber \\
    \left.
    {}+ \varepsilon_{\rm h1}B_n^{+}C_{n-1,1} + \varepsilon_{\rm h2}B_n^{+}C_{n+1,2} + \mbox{h.c.} \right],
\end{eqnarray}
where $\varepsilon_{\rm e1}$ and $\varepsilon_{\rm e2}$ are the transfer integrals of the electron from molecule $n$ to molecules $n+1$ and $n-1$ correspondingly, and $\varepsilon_{\rm h1}$, $\varepsilon_{\rm h2}$ denote the transfer integrals of the hole from molecule $n$ to molecules $n+1$ and $n-1$.  Certainly the model with four transfer integrals is more complicated than the model in Refs.~\cite{hoffmann00,hoffmann02} with two mixing parameters only.  But our model is more realistic because we take into account the inclination of the flat molecules of PTCDA and MePTCDI relatively to the stack axis (see the Appendix).

One intramolecular mode is only supposed to be coupled with the FE and CTEs and the phonon part of the Hamiltonian is
\begin{equation}
\label{eq:phonhamilt}
    \hat{H}_{\text{ph}} = \sum_n \hbar \omega_0 a_n^{+} a_n,
\end{equation}
where $\omega_0$ is the vibrational frequency and $a_n$ is the annihilation operator of one vibrational quantum on molecule $n$.  The linear exciton--phonon coupling only is manifested in the treated crystals \cite{hoffmann02} and we get the following exciton--phonon part \cite{henessy99,hoffmann00,hoffmann02,lalov05}
\begin{eqnarray}
\label{eq:exc-phon}
    \hat{H}_{\text{ex--phon}} = \sum_{n,\sigma=1,2}\hbar \omega_0\left[ \xi_{\rm F}B_n^{+} B_n \left( a_n^{+} + a_n \right)\right. \nonumber \\
    \left. {}+ \xi C_{n\sigma}^{+} C_{n\sigma} \left( a_n^{+} + a_n + a_{n+\sigma_1}^{+} + a_{n+\sigma_1} \right) \right],
\end{eqnarray}
where $\xi_{\rm F}$ and $\xi$ are dimensionless parameters of the linear FE--phonon and CTEs--phonon coupling, correspondingly; $\sigma_1 = +1$ if $\sigma = 1$, and $\sigma_1 = -1$ if $\sigma = 2$.  For the sake of simplicity, we suppose the same linear exciton--phonon coupling in the positive and negative ions which create CTEs.

The full Hamiltonian $\hat{H}$ contains all the parts (\ref{eq:fehamilt})--(\ref{eq:exc-phon}) and can be transformed using the canonical transformation which eliminates the linear exciton--phonon coupling (\ref{eq:exc-phon}), see \cite{lalov05,davydov71},
\begin{equation}
\label{eq:h1}
    \hat{H}_1 = \exp(Q) \hat{H} \exp(-Q),
\end{equation}
where
\begin{eqnarray}
\label{eq:q}
    Q = \sum_{n,\sigma=1,2} \left[ \xi_{\rm F}B_n^{+} B_n \left( a_n^{+} - a_n \right) \right. \nonumber \\
    \left. {}+ \xi C_{n\sigma}^{+} C_{n\sigma} \left( a_n^{+} - a_n + a_{n+\sigma_1}^{+} - a_{n+\sigma_1} \right) \right].
\end{eqnarray}

We introduce the vibronic operators
\begin{equation}
\label{eq:vn}
    V_n = \exp(Q) B_n \exp(-Q),
\end{equation}
\begin{equation}
\label{eq:unsigma}
    U_{n,\sigma} = \exp(Q) C_{n\sigma} \exp(-Q)
\end{equation}
and get the following transformed Hamiltonian
\begin{eqnarray}
\label{eq:h1new}
    \hat{H}_1 = \sum_n \left( E_{\rm F} - \hbar \omega_0 \xi_{\rm F}^2 \right)V_n^{+}V_n + \sum_{n n^{\prime}} L \left( \delta_{n^{\prime},n+1} + \delta_{n^{\prime},n-1} \right) V_{n^{\prime}}^{+} V_n \nonumber \\
    {}+ \sum_{n,\sigma} \left( E_{\rm c} - 2\hbar \omega_0 \xi^2 \right)U_{n \sigma}^{+}U_{n \sigma} + \sum_n \hbar \omega_0 a_n^{+} a_n + \sum_n \left[ \varepsilon_{\rm e1}V_n^{+}U_{n1} \right. \nonumber \\
    \left.
    {}+ \varepsilon_{\rm e2}V_n^{+}U_{n2} + \varepsilon_{\rm h1}V_n^{+}U_{n-1,1} + \varepsilon_{\rm h2}V_n^{+}U_{n+1,2} + \mbox{h.c.} \right]
\end{eqnarray}

In a stack with inversion center, point group $C_i$, the excitons are \emph{gerade\/} or \emph{ungerade\/} (in the center of the Brillouin zone, at $k = 0$).  The ungerade excitons only are dipole-active and influence the linear optical susceptibility and the absorption spectra.  In the case under consideration, the operator of the transition dipole moment has the following form \cite{lalov08,haarer75}:
\begin{equation}
\label{eq:p}
    P = \sum_n \left[ \mathbf{p}_{\rm F}\left( V_n^{+} + V_n \right) + \mathbf{p}_{\rm CT}\left( U_{n2} - U_{n1} + U_{n2}^{+} - U_{n1}^{+} \right)  \right].
\end{equation}

The gerade FEs, as well as the symmetrical combination of CTEs, $\left( U_{n2}^{+}\right.$ $\left.+\; U_{n1}^{+} \right)|0\rangle$, can be also mixed, but due to their vanishing transition dipole moment they will not be considered here.

Introducing the Fourier transform in the momentum space of the vibronic operators,
\begin{equation}
\label{eq:vk}
    V_k = \frac{1}{\sqrt{N}}\sum_n V_n \exp(\mathrm{i}knd),
\end{equation}
\begin{equation}
\label{eq:uksigma}
    U_{k,\sigma} = \frac{1}{\sqrt{N}}\sum_n U_{n,\sigma} \exp(\mathrm{i}knd),
\end{equation}
we obtain the following form of the Hamiltonian (\ref{eq:h1new})
\begin{eqnarray}
\label{eq:h1new1}
    \hat{H}_1 \!\!\!\!\!&=&\!\!\!\!\! \sum_k \left( E_{\rm F} - \hbar \omega_0 \xi_{\rm F}^2 + 2L\cos k \right)V_k^{+}V_k \nonumber \\
    &&\!\!\!\!\!
    {}+ \sum_{k,\sigma} \left( E_{\rm c} - 2\hbar \omega_0 \xi^2 \right) U_{k\sigma}^{+}U_{k\sigma} \nonumber \\
    &&\!\!\!\!\!
    {}+ S + \sum_n \hbar \omega_0 a_n^{+} a_n,
\end{eqnarray}
where
\[
    S = \sum_k \left\{ A V_k^{+}\left( U_{k,1} + U_{k,2} \right) + B V_k^{+}\left( U_{k,2} - U_{k,1} \right) + \mbox{h.c.} \right\}
\]
with
\[
    A = \varepsilon_{\rm e}^{\prime} + \varepsilon_{\rm h}^{\prime}\cos k + \mathrm{i}\varepsilon_{\rm h}^{\prime \prime} \sin k \quad \mbox{and} \quad
    B =  \varepsilon_{\rm e}^{\prime \prime} + \varepsilon_{\rm h}^{\prime \prime}\cos k
    + \mathrm{i}\varepsilon_{\rm h}^{\prime} \sin k .
\]
Here
\begin{equation}
\label{eq:epsprime}
    \varepsilon_{\rm e}^{\prime} = \left( \varepsilon_{\rm e1} + \varepsilon_{\rm e2} \right)/2, \qquad \varepsilon_{\rm h}^{\prime} = \left( \varepsilon_{\rm h1} + \varepsilon_{\rm h2} \right)/2,
\end{equation}
\begin{equation}
\label{eq:epssec}
    \varepsilon_{\rm e}^{\prime \prime} = \left( \varepsilon_{\rm e2} - \varepsilon_{\rm e1} \right)/2, \qquad \varepsilon_{\rm h}^{\prime \prime} = \left( \varepsilon_{\rm h2} - \varepsilon_{\rm h1} \right)/2.
\end{equation}
For the case of ungerade FEs, their mixing with the symmetrical (even) combination $\left( U_{k,1} + U_{k,2} \right)$ at $k = 0$ is impossible, and thus the mixing parameters $\varepsilon_{\rm e}^{\prime}$ and $\varepsilon_{\rm h}^{\prime}$ vanish, $\varepsilon_{\rm e}^{\prime} = \varepsilon_{\rm h}^{\prime} = 0$.  The final expressions for operators (\ref{eq:fcoperator}) and (\ref{eq:p}) are
\begin{eqnarray}
\label{eq:fc}
    \hat{H}_{\rm FCT} = \sum_k \left[ \left( \varepsilon_{\rm e}^{\prime \prime} + \varepsilon_{\rm h}^{\prime \prime}\cos k \right) V_k^{+} \left( U_{k,2} - U_{k,1} \right) \right. \nonumber \\
    \left.
    {}+ \mathrm{i}\varepsilon_{\rm h}^{\prime \prime}\sin k V_k^{+} \left( U_{k,2} + U_{k,1} \right) + \mbox{h.c.} \right]
\end{eqnarray}
and
\begin{eqnarray}
\label{eq:pfinal}
    \hat{P} = \sqrt{N}\left[ \mathbf{p}_{\rm F}\left( V_{k=0} + V_{k=0}^{+} \right) \right. \nonumber \\
    \left.
    {}+ \mathbf{p}_{\rm CT}\left( U_{k=0,2} - U_{k=0,1} + U_{k=0,2}^{+} - U_{k=0,1}^{+} \right) \right],
\end{eqnarray}
respectively.

\section{Calculation of the Linear Optical Susceptibility}
\label{sec:optical}

The linear optical susceptibility can be calculated by using the formula \cite{agranovich83}
\begin{equation}
\label{eq:chiij}
    \chi_{ij} = \lim_{\epsilon \to 0}\left\{ \frac{1}{2\hbar V}\left[ \Phi_{ij}(\omega + \mathrm{i}\epsilon) + \Phi_{ij}(-\omega + \mathrm{i}\epsilon) \right] \right\}
\end{equation}
with
\begin{equation}
\label{eq:phiij}
    \Phi_{ij}(t) = -\mathrm{i} \theta(t)\langle 0| \hat{P}_i(t)\hat{P}_j(0) + \hat{P}_j(t)\hat{P}_i(0)|0\rangle,
\end{equation}
where $V$ is the crystal's volume [in our case being proportional to $Nv$ ($v$ is the volume occupied by one molecule)] and $\hat{P}$ is the operator (\ref{eq:pfinal}).  The Green functions (\ref{eq:phiij}) have been calculated as average over the ground state $|0\rangle$ only by taking into account the large values of $E_{\rm F}$, $E_{\rm c}$, $\hbar \omega_0 \gg k_{\rm B}T$.

We calculate the Green functions (\ref{eq:phiij}) following the vibronic approach \cite{lalov07b,lalov08}.  In the next expression, the $x$ axis is supposed to be oriented along the vector $\mathbf{p}_{\rm F}$ which includes angle $\gamma$ with the vector $\mathbf{p}_{\rm CT}$, and $a = \left| p_{\rm CT}/p_{\rm F} \right|$.  Then we can represent the linear optical susceptibility of one stack as
\begin{equation}
\label{eq:chixx}
    \chi_{xx} = -\frac{p_{\rm F}^2}{v}\frac{1}{\alpha_1 \alpha_2 - 2\alpha_{12}^2} \left[ \alpha_2 + 4a\alpha_{12}\cos \gamma + 2a^2 \alpha_1 \cos^2 \gamma \right].
\end{equation}
The functions $\alpha_1$, $\alpha_2$, $\alpha_{12}$ have been calculated for the excitonic and one-phonon vibronic regions (see below).  The PTCDA and MePTCDI crystals contain two types $A$ and $B$ of parallel molecular stacks, however, the excitonic and vibronic excitations in each stack interact very weakly with the excitations of the other stacks.  In the same way as in Ref.~\cite{lalov06b}, we calculate the crystal's susceptibility in an oriented gas model.  We denote by $2\varphi$ the angle between the vectors $\mathbf{p}_{\rm F}^A$ and $\mathbf{p}_{\rm F}^B$ of two different stacks and suppose that these vectors are positioned in the $(XY)$ plane, the crystal $X$ axis been oriented along the sum $\mathbf{p}_{\rm F}^A + \mathbf{p}_{\rm F}^B$ \cite{note}.

The components of the linear optical susceptibility of the crystal correspondingly are:
\begin{eqnarray}
\label{eq:chixxnew}
    \chi_{XX} = -\frac{p_{\rm F}^2}{v}\frac{2}{\alpha_1 \alpha_2 - 2\alpha_{12}^2} \left[ \alpha_2 \cos^2 \varphi \right. \nonumber \\
    \left.
    {}+ 4a\alpha_{12}\cos^2 \varphi \cos \gamma + a^2 \alpha_1 (1 + \cos 2\gamma \cos \varphi) \right]
\end{eqnarray}
and
\begin{eqnarray}
\label{eq:chiyy}
    \chi_{YY} = -\frac{p_{\rm F}^2}{v}\frac{2}{\alpha_1 \alpha_2 - 2\alpha_{12}^2} \left[ \alpha_2 \sin^2 \varphi \right. \nonumber \\
    \left.
    {}+ 4a\alpha_{12}\sin^2 \varphi \cos \gamma + a^2 \alpha_1 (1 - \cos 2\gamma \cos \varphi) \right].
\end{eqnarray}

We find the following expressions for functions $\alpha_1$, $\alpha_2$, $\alpha_{12}$:

(1) In the excitonic region expressions practically coincide with formulas in Ref.~\cite{lalov08}, namely
\begin{equation}
\label{eq:alpha1}
    \alpha_1 = \hbar \omega - \left( E_{\rm F} + 2L \right) - \hbar \Omega_{\rm 0F}(1),
\end{equation}
\begin{equation}
\label{eq:alpha12}
    \alpha_{12} = \varepsilon_{\rm e}^{\prime \prime} + \varepsilon_{\rm h}^{\prime \prime},
\end{equation}
\begin{equation}
\label{eq:alpha2}
    \alpha_2 = \hbar \left[ \omega - \Omega_{\rm 0c}(1) \right] - E_{\rm c},
\end{equation}
where $\Omega_{\rm 0F}(1)$ and $\Omega_{\rm 0c}(1)$ are expressed through the continuous fractions following from recursions:
\begin{equation}
\label{eq:omega0f}
    \Omega_{\rm 0F}(n) = \frac{n\omega_{\rm a}^2}{\omega - \left( E_{\rm F} + 2L \right)/\hbar - n\omega_0 - \Omega_{\rm 0F}(n+1)},
\end{equation}
\begin{equation}
\label{eq:omega0c}
    \Omega_{\rm 0c}(n) = \frac{2n\omega_{\rm a1}^2}{\omega - E_{\rm c}/\hbar - n\omega_0 - \Omega_{\rm 0c}(n+1)},
\end{equation}
\begin{equation}
\label{eq:omegaa}
    \omega_{\rm a} = \xi_{\rm F}\omega_0, \qquad \omega_{\rm a1} = \xi \omega_0.
\end{equation}

(2) In the one-phonon vibronic region
\begin{equation}
\label{eq:alfa1}
    \alpha_1 = \hbar \omega - \left( E_{\rm F} + 2L \right) - \frac{\omega_{\rm a}^2}{mD} \left[ (\hbar \omega_{\rm 1c} - \beta)\sigma - \frac{\beta \left( \varepsilon_{\rm e}^{\prime \prime} + \varepsilon_{\rm h}^{\prime \prime} \right)^2}{m}\sigma_1 \right],
\end{equation}
\begin{equation}
\label{eq:alfa12}
    \alpha_{12} = \left( \varepsilon_{\rm e}^{\prime \prime} + \varepsilon_{\rm h}^{\prime \prime} \right)\left[ 1 + \frac{\omega_{\rm a} \omega_{\rm a1}}{mD}\left( \sigma + \sigma_1 \right) \right],
\end{equation}
\begin{eqnarray}
\label{eq:alfa2}
    \alpha_2 = \hbar \omega - E_{\rm c} - \frac{2\omega_{\rm a1}^2}{mD} \left\{ \frac{m}{\omega_{\rm 1c}} - \alpha_{\rm F}\sigma + \left( \varepsilon_{\rm e}^{\prime \prime} + \varepsilon_{\rm h}^{\prime \prime} \right)^2 \right. \nonumber \\
    \left. \times \! \left[ \frac{\sigma + \sigma_1 (2 + t)}{\omega_{\rm 1c}} - \frac{\alpha_{\rm F}\sigma_1}{m} \right] \right\},
\end{eqnarray}
where
\begin{equation}
\label{eq:omega1c}
    \omega_{\rm 1c} = \omega - \omega_0 - \Omega_{\rm 1c}(1) - E_{\rm c}/\hbar,
\end{equation}
\vspace{1mm}
\begin{equation}
\label{eq:m}
    m = 2\left[ L\hbar \omega_{\rm 1c} + 2 \varepsilon_{\rm e}^{\prime \prime} \varepsilon_{\rm h}^{\prime \prime} \right],
\end{equation}
\begin{eqnarray}
\label{eq:t}
    t = (1/m)\left\{
    \left\{ \hbar \left[ \omega - \omega_0 - \Omega_{\rm 1F}(1) \right]- E_{\rm F} \right\} \hbar \omega_{\rm 1c}
    - 2\left[ \left( \varepsilon_{\rm e}^{\prime \prime} \right)^2 + \left( \varepsilon_{\rm h}^{\prime \prime} \right)^2 \right] \right\},
\end{eqnarray}
\begin{eqnarray}
\label{eq:D}
    D = \left( 1 - \frac{\alpha_{\rm F} \omega_{\rm 1c} \sigma}{m} \right) \left( 1 - \frac{\beta}{\omega_{\rm 1c}} \right) \nonumber \\
    {}+ \frac{\beta \left( \varepsilon_{\rm e}^{\prime \prime} + \varepsilon_{\rm h}^{\prime \prime} \right)^2}{m} \left[ \frac{\alpha_{\rm F} \sigma_1}{m} - \frac{\sigma + \sigma_1(2 + t)}{\omega_{\rm 1c}} \right],
\end{eqnarray}
\vspace{2mm}
\begin{equation}
\label{eq:sigma}
    \sigma = \left\{ \begin{array}{cc}
    -1/\sqrt{t^2 - 1} & \mbox{if\,\,\, $\left| t + \sqrt{t^2 - 1} \right| < 1$,}
    \\ \\
    1/\sqrt{t^2 - 1} & \mbox{if\,\,\, $\left| t + \sqrt{t^2 - 1} \right| < 1$,}
                     \end{array}
    \right.
\end{equation}
and
\begin{equation}
\label{eq:sigma1}
    \sigma_1 = t\sigma -1.
\end{equation}

The functions $\Omega_{\rm 1F}(1)$ and $\Omega_{\rm 1c}(1)$ also represent continuous fractions from recursions:
\begin{equation}
\label{eq:omega1f}
    \Omega_{\rm 1F}(n) = \frac{n\omega_{\rm a}^2}{\omega - E_{\rm F}/\hbar - (n+1)\omega_0 - \Omega_{\rm 1F}(n+1)},
\end{equation}
\begin{equation}
\label{eq:Omega1c}
    \Omega_{\rm 1c}(n) = \frac{2n\omega_{\rm a1}^2}{\omega - E_{\rm c}/\hbar - (n+1)\omega_0 - \Omega_{\rm 1c}(n+1)}.
\end{equation}
Finally, the functions $\beta$ and $\alpha_{\rm F}$ can be expressed as follows:
\begin{equation}
\label{eq:beta/h}
    \beta/\hbar = \Omega_{\rm 0c}(2) - \Omega_{\rm 1c}(1),
\end{equation}
\begin{equation}
\label{eq:alphaf/h}
    \alpha_{\rm F}/\hbar = \Omega_{\rm 0F}(2) - \Omega_{\rm 1F}(1).
\end{equation}

\section{Simulations of the Excitonic and Vibronic Spectra of MePTCDI and PTCDA Crystals}
\label{sec:simul}
In this section, we calculate the absorption spectra of the two crystals finding the imaginary parts of the components of the linear optical susceptibility (\ref{eq:chixxnew}) at $\left( 2p_{\rm F}^2/v \right) = 1$ and supposing an imaginary part equal to $\mathrm{i}\delta/\hbar$ of the frequency $\omega$.  We put the excitonic and vibrational parameters for the studied crystals as they have been fitted in Refs.~\cite{henessy99} and \cite{schmidt02} and used in our previous papers \cite{lalov06a,lalov07a,lalov06b,lalov06c}, see table 1:
\begin{table}
\caption{\label{data}Excitonic and vibrational parameters (in cm$^{-1}$) of the MePTCDI and PTCDA crystals.}{\smallskip}
\begin{small}\centering
\begin{tabular}{lccccrccc}
\hline \noalign {\smallskip}
 & $E_{\rm F}$ & $E_{\rm c}$ & $\hbar \omega_0$ & $L$ & $\varepsilon_{\rm e}$ & $\varepsilon_{\rm h}$ & $2\varphi$ & $\gamma$ \\
\hline \noalign
{\smallskip}
MePTCDI & $17\,992$ & $17\,346$ & $1\,400$ & $345$ & $-380$ & $-137$ & $36.8^\circ$ & $68.1^\circ$ \\
PTCDA   & $18\,860$ & $18\,300$ & $1\,400$ & $330$ & $-48$ & $-436$ & $82^\circ$ & $143^\circ$ \\
\hline
\end{tabular}
\end{small}
\end{table}
The data for angle $\gamma$ and ratio $a = \left| p_{\rm CT}/p_{\rm F} \right|$ have been calculated in \cite{hoffmann00} using quantum chemical evaluations.  We use the values $a = 0.1$ for the PTCDA crystal and $a = 0.135$ for the MePTCDI.  The data for angle $2\varphi$ are derived from the crystal structure (for MePTCDI see \cite{haedicke86}).

The aforementioned parameters are permanent in our calculations.  We vary the values of the following parameters (intending to observe their impact on the absorption spectra and find a better similarity with the experimental absorption spectra \cite{henessy99,hoffmann00,hoffmann02}):

(i) The excitonic damping quantity $\delta$ which varies from $1$ to $500$ cm$^{-1}$.

(ii) The linear exciton--phonon coupling parameters $\xi_{\rm F}$ and $\xi$ which vary near the values
\begin{eqnarray*}
    \xi_{\rm F} \!\!\!\!\!&=&\!\!\!\!\! 0.82 \mbox{ and } 1.1 \quad \mbox{for PTCDA} \quad \mbox{(see \cite{henessy99})},
    \\
    \xi_{\rm F} \!\!\!\!\!&=&\!\!\!\!\! 0.88 \mbox{ and } 1.1 \quad \mbox{for MePTCDI} \quad \mbox{(see \cite{hoffmann02})},
\end{eqnarray*}
and
\[
    \xi = \xi_{\rm F}, \mbox{ }\xi_{\rm F}/\sqrt{2}, \mbox{ and }\; \xi_{\rm F}/2 \;\, \mbox{for both crystals}.
\]
In calculating continuous fractions (\ref{eq:omega0f}), (\ref{eq:omega0c}) and (\ref{eq:omega1f}), (\ref{eq:Omega1c}) we take twenty steps.

\subsection{MePTCDI}
\label{subsec:meptcdi}
The fine structure of the excitonic spectra of this crystal is presented in Figure~1.  The absorption curves are calculated at small value $\delta = 20$ cm$^{-1}$ of the excitonic damping for $\xi_{\rm F} = 0.88$ with $\xi = \xi_{\rm F}$ (blue curve 1), $\xi = \xi_{\rm F}/\sqrt{2}$ (red curve 2) and $\xi = \xi_{\rm F}/2$ (green curve 3).  We have calculated the imaginary part of expression~(\ref{eq:chixxnew}) using the `excitonic' formulas (\ref{eq:alpha1})--(\ref{eq:alpha2}) (in the following denoted as ``exc'' program).  For the smaller values of $\xi$ equal to $\xi_{\rm F}/\sqrt{2}$ and $\xi_{\rm F}/2$ we obtain spectral doublets which correspond to the CTEs--FE splitting.  This splitting generates a spectral triplet at $15\,100$ cm$^{-1}$ (very weak), $16\,500$ cm$^{-1}$, and $17\,200$ cm$^{-1}$ (for $\xi = \xi_{\rm F}$, blue curve 1).
\begin{figure}[!ht]
\centering\includegraphics[height=.25\textheight]{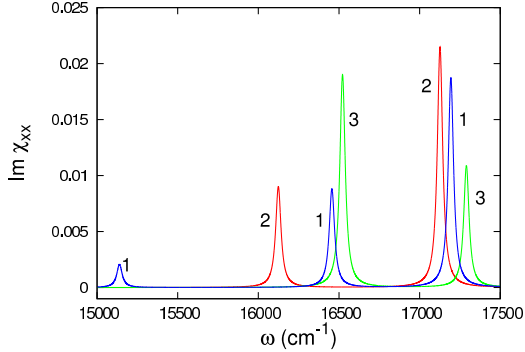}
  \caption{Linear absorption in the excitonic region of the MePTCDI crystal at $\delta = 20$ cm$^{-1}$, $\xi_{\rm F} = 0.88$ with $\xi = \xi_{\rm F}$ (blue curve 1), $\xi = \xi_{\rm F}/\sqrt{2}$ (red curve 2) and $\xi = \xi_{\rm F}/2$ (green curve 3) calculated using formulas (24)--(26) (``exc'' program).}
  \label{fig:fg1}
\end{figure}
\begin{figure}[!ht]
\centering\includegraphics[height=.25\textheight]{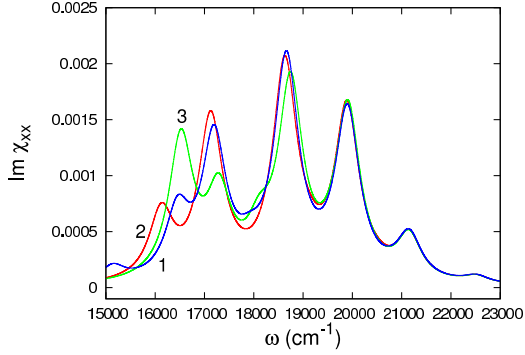}
  \caption{Linear absorption in the excitonic region of the MePTCDI crystal at $\delta = 300$ cm$^{-1}$ (``exc'' program).  For the values of $\xi_{\rm F}$ and $\xi$ see Figure~\ref{fig:fg1}.}
  \label{fig:fg2}
\end{figure}
\begin{figure}[!ht]
\centering\includegraphics[height=.25\textheight]{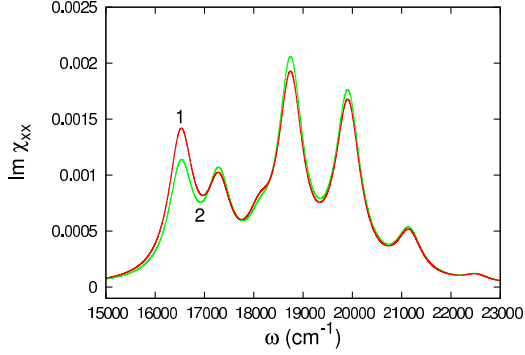}
  \caption{Impact of the CTEs-transition dipole moment expressed through the ratio $a$.  Red curve 1 for $a = 0.135$ and the green curve 2 for $a = 0$ ($\delta = 300$ cm$^{-1}$, $\xi_{\rm F} = 0.88$, $\xi = \xi_{\rm F}/2$).}
  \label{fig:fg3}
\end{figure}

The value $\delta = 300$ cm$^{-1}$ seems to be more closer to the width of the lowest absorption maximum observed in the MePTCDI crystal \cite{hoffmann00,hoffmann02}.  The absorption curves in Figure~2 are calculated at $\delta = 300$ cm$^{-1}$ in the excitonic region (approximately below $17\,800$ cm$^{-1}$), one-phonon vibronic region ($17\,800$--$19\,200$ cm$^{-1}$), two-phonon vibronic region ($19\,200$--$20\,700$ cm$^{-1}$) etc.  There exists an obvious similarity between experimental curve (see Ref.~\cite{hoffmann00}) and the green curve 3 in Figure~2 calculated with $\xi = \xi_{\rm F}/2$.

The impact of the CTEs transition dipole moment is illustrated in Figure~3.  The two curves -- the red one 1 calculated with the CTEs contribution, and the green one 2 calculated without this contribution [$a \equiv 0$ in formulas (\ref{eq:chixx})--(\ref{eq:chiyy})] -- are relatively close to each other.  The strongest impact can be measured near lower excitonic peak generated by the CTEs level $E_{\rm c} = 17\,346$ cm$^{-1}$.
\begin{figure}[!ht]
\centering\includegraphics[height=.25\textheight]{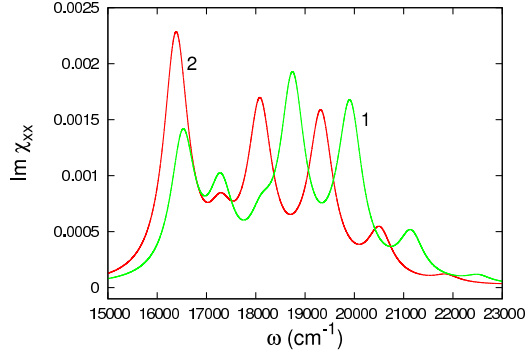}
  \caption{Comparison of the absorption curves calculated with the data from Table~1 (green curve 1) and the red curve 2 calculated for $E_{\rm F} = 17\,346$ cm$^{-1}$ and $E_{\rm c} = 17\,992$ cm$^{-1}$ (interchange of the excitonic levels) with $\delta = 300$ cm$^{-1}$, $\xi_{\rm F} = 0.88$, $\xi = \xi_{\rm F}/2$.}
  \label{fig:fg4}
\end{figure}

The two curves in Figure~4 calculated by using the ``exc'' program for $\delta = 300$ cm$^{-1}$ differ by the mutual positions of the two excitonic levels. The green curve 1 corresponds to a lower position of the CTEs level $E_{\rm c} < E_{\rm F}$, whereas the red curve 2 corresponds to the opposite situation $E_{\rm F} < E_{\rm c}$ (the magnitudes of $E_{\rm c}$ and $E_{\rm F}$ are exchanged).  The red curve 2 exhibits a strong domination of the lower maxima associated with the FE level. The two curves can be approximated with five Lorentz maxima in the spectral region of $15\,000$--$22\,000$ cm$^{-1}$ \cite{hoffmann00}.  Comparing the lineshapes of these two curves with the experimental curve in Ref.~\cite{hoffmann00}, we cannot make a hypothesis which possibility is more probable.  We prefer the fitting from Refs.~\cite{hoffmann00} and \cite{schmidt02} $\left( E_{\rm c} < E_{\rm F} \right)$, however, another choice, $ E_{\rm F} < E_{\rm c}$, is also allowed.
\begin{figure}[!ht]
\centering\includegraphics[height=.25\textheight]{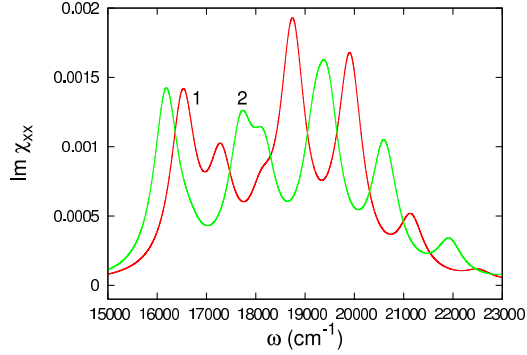}
  \caption{The impact of the FE--phonon coupling constant $\xi_{\rm F}$ on the linear absorption.  The red curve 1 corresponds to $\xi_{\rm F} = 0.88$, and the green curve 2 to $\xi_{\rm F} = 1.1$ ($\xi = \xi_{\rm F}/2$, $\delta = 300$ cm$^{-1}$).}
  \label{fig:fg5}
\end{figure}
\begin{figure}[!ht]
\centering\includegraphics[height=.25\textheight]{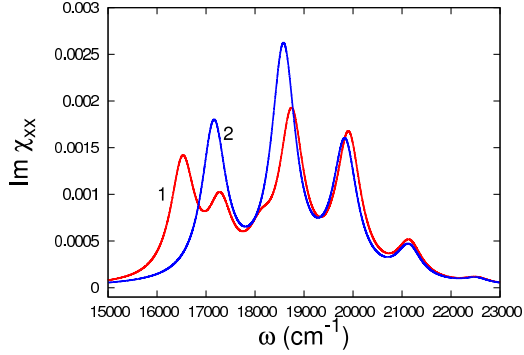}
  \caption{Linear absorption spectra of the MePTCDI crystal with FE--CTEs mixing (red curve 1) and pure Frenkel-exciton spectra (blue curve 2 at $\varepsilon_{\rm e}^{\prime \prime} = \varepsilon_{\rm h}^{\prime \prime} = 0$) obtained by the ``exc'' program with $\delta = 300$ cm$^{-1}$, $\xi_{\rm F} = 0.88$, $\xi = \xi_{\rm F}/2$.}
  \label{fig:fg6}
\end{figure}

Figure~5 illustrates the impact of the FE--phonon coupling constant $\xi_{\rm F}$ on the absorption spectra.  The values $\xi_{\rm F} = 0.88$ and $\xi = \xi_{\rm F}/2$ can be considered as better candidates for the simulation (compared with the experimental absorption lineshape especially in the excitonic region).

The importance of the FE--CTEs mixing can be seen in Figure~6 in which the blue curve 2 represents a pure FE-absorption $\left( \varepsilon_{\rm e}^{\prime \prime} = \varepsilon_{\rm h}^{\prime \prime} = 0 \right)$.  The complexity of the experimental absorption curves with five Lorentz maxima in the studied spectral region, see Figures~2 and 3 in Ref.~\cite{hoffmann00}, cannot be understood on the basis of simple Frenkel-exciton model.  Contrary, the model of mixed FE--CTEs and their vibronic satellites can be the basis of adequate simulations of the absorption curves (red curve 1).
\begin{figure}[!ht]
\centering\includegraphics[height=.25\textheight]{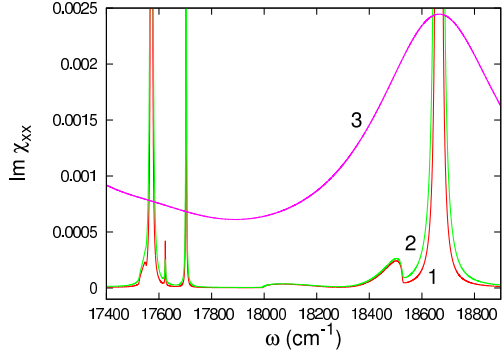}
  \caption{One-phonon vibronic spectra (``1p'' program) of the MePTCDI crystal with $\xi_{\rm F} = 0.88$, $\xi = \xi_{\rm F}/\sqrt{2}$, $\delta = 1$ cm$^{-1}$ (red curve 1), $\delta = 2$ cm$^{-1}$ (green curve 2), and $\delta = 300$ cm$^{-1}$ (purple curve 3).}
  \label{fig:fg7}
\end{figure}

The absorption curves in the following three Figures~7--9 have been calculated by using formulas (\ref{eq:omega0f})--(\ref{eq:alphaf/h}) for one-phonon vibronic spectra denoted in the following as ``1p'' program.  Our calculations based on the Green functions formalism allow to study and simulate the absorption associated with the two types of exciton--phonon states in the vibronic spectra, notably:  (a) bound states (one-particle states) corresponding to the propagation of excitons and phonon(s) in the stack as a whole.  The lineshape of their absorption maxima is Lorentzian and strongly depends on the excitonic damping $\delta$. (b) Many-particle states (MP) corresponding to the excitation of exciton and phonon(s) on separate molecules of the stack.  Their energy lies in the quasicontinuous band(s) and the lineshape of the corresponding absorption maxima depends on the density of the states in the MP band and more weakly on the excitonic damping $\delta$.  Unfortunately the big values of $\delta$ in both crystals under consideration mask the MP bands which are totally covered by the absorption maxima.  We can simulate the absorption in MP continua by supposing very small values of the damping, for instance, $\delta \sim 1$--$10$ cm$^{-1}$, which is not typical for the perylene derivatives (but occurring in some DA-crystals like An-PMDA \cite{sebastian81,sebastian83,haarer75,brillante80}).
\begin{figure}[ht]
  \begin{minipage}[b]{0.5\linewidth}
  \centering
    \includegraphics[width=5.5cm]{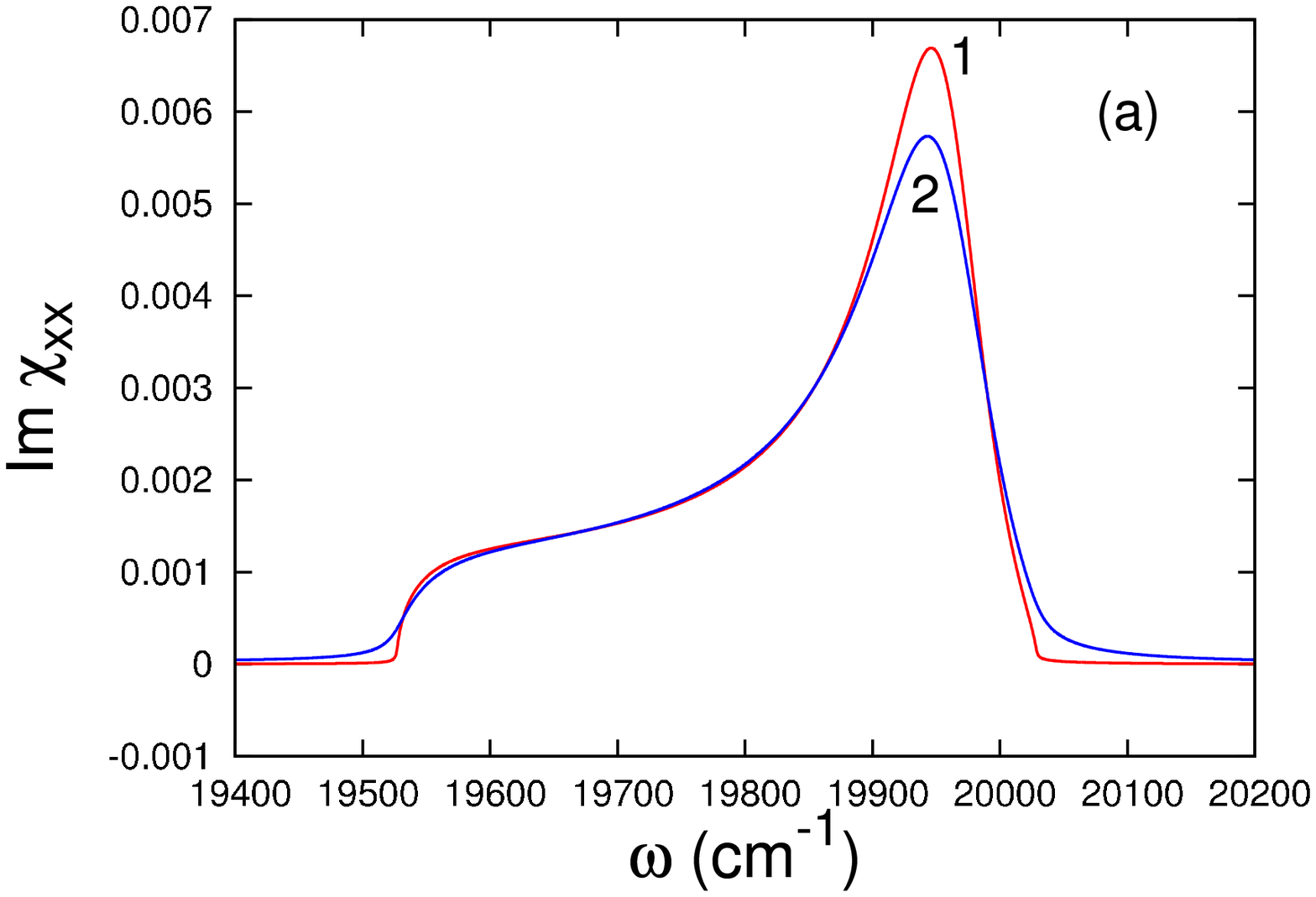}
  \end{minipage}
  \hspace{-0.2cm}
  \begin{minipage}[b]{0.5\linewidth}
  \centering
    \includegraphics[width=5.5cm]{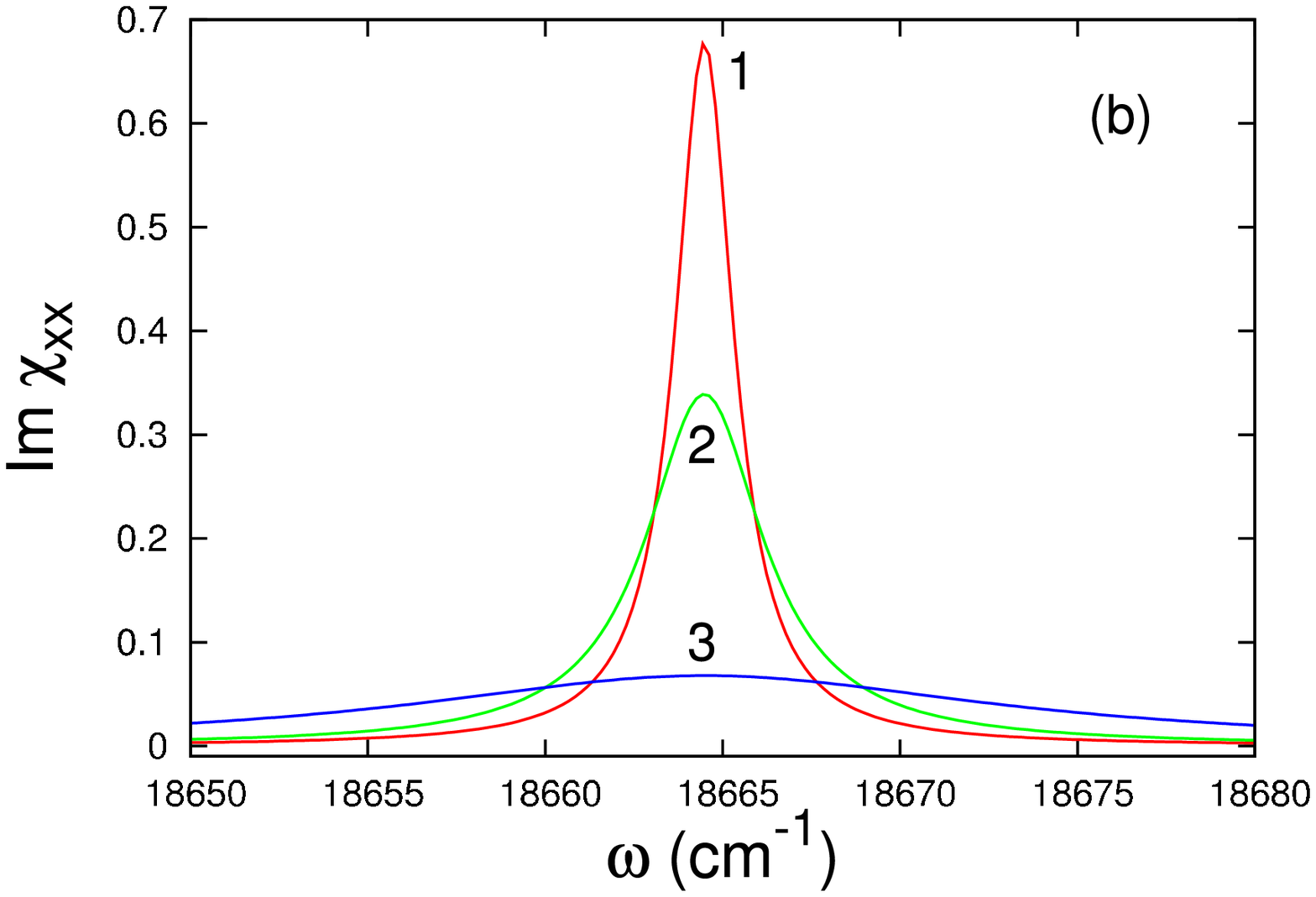}
  \end{minipage}
  \caption{(a) MP band in two-phonon vibronic spectra with $\xi_{\rm F} = 0.88$, $\xi = \xi_{\rm F}/\sqrt{2}$, $\delta = 1$ cm$^{-1}$ (red curve 1), $\delta = 10$ cm$^{-1}$ (blue curve 2). (b) Bound exciton--phonon state in one-phonon vibronic spectra with $\xi_{\rm F} = 0.88$, $\xi = \xi_{\rm F}/\sqrt{2}$, $\delta = 1$ cm$^{-1}$ (red curve 1), $\delta = 2$ cm$^{-1}$ (green curve 2), $\delta = 10$ cm$^{-1}$ (blue curve 3).}
  \label{fig:fg8}
\end{figure}
\begin{figure}[!ht]
\centering\includegraphics[height=.25\textheight]{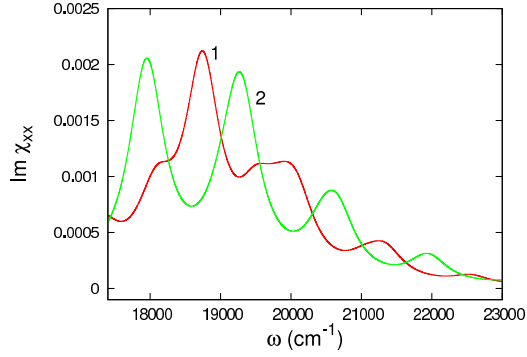}
  \caption{Linear absorption in the vibronic spectra of the MePTCDI crystal (``1p'' program) with FE--CTEs mixing (red curve 1) and pure FE vibronics  (green curve 2 at $\varepsilon_{\rm e}^{\prime \prime} = \varepsilon_{\rm h}^{\prime \prime} = 0$) with $\delta = 300$ cm$^{-1}$, $\xi_{\rm F} = 0.88$, $\xi = \xi_{\rm F}/2$.}
  \label{fig:fg9}
\end{figure}

The one-phonon vibronic spectra are depicted in Figure~7.  The vibronics of FE and CTEs are mixed like their excitons' spectra.  Nevertheless, the less intensive lower maximum at $17\,500$--$17\,700$ cm$^{-1}$ can be considered as vibronic replica of CTEs whereas the more intensive maximum near $18\,400$--$18\,700$ cm$^{-1}$ is the replica of the FE.  The lower wings of both maxima represent the very weak MP bands (even virtual ones).  The main part of absorption intensity is concentrated above those MP bands in the Lorentzian maxima near $17\,600$ cm$^{-1}$ and $18\,650$ cm$^{-1}$.  As can be seen in Figure~7, the wide absorption maxima corresponding to the realistic value $\delta = 300$ cm$^{-1}$ (purple curve 3) cover the fine structure of the absorption lineshape.

Figure~8 illustrates the difference between the MP band (left panel (a)) and the bound exciton--phonon maximum (right panel (b)).  The line shape of the MP band depends of the density of the states which is strongly modulated by the exciton--phonon coupling.  Thus, the lineshape is non-symmetrical.  In the case of relatively small damping parameter $\delta$ the lineshape (look at Figure~8(a)) depends very weakly on $\delta$.  Contrary, the maximal value of the Lorentzian maximum in Figure~8(b) is proportional to $1/\delta$.

Figure~9 is analog of Figure~6 in the vibronic region (above $17\,500$ cm$^{-1}$).  The lineshape of the red curve 1 (with FE--CTEs mixing) is more complicated than the lineshape of the pure FE vibronics.  The red curve 1 seems to be a better simulation of the experimental absorption curve \cite{hoffmann00}.

The model of the FE--CTEs mixing reproduces the general structure of the absorption spectra in the excitonic and vibronic regions of the MePTCDI crystal (see Figures~2 and 9).  Our calculations confirm the correct choice of the excitonic and vibrational parameters in the fitting procedure implemented in Refs.~\cite{hoffmann00,schmidt02}, especially the value $\xi_{\rm F} = 0.88$ of the constant of the linear FE--phonon coupling.  Additionally, we establish as the most probable value of the exciton damping $\delta = 300$ cm$^{-1}$ and CTEs--phonon coupling constant $\xi = \xi_{\rm F}/2$.  Our studies show the positions of the bands both of FE$+$phonon and CTEs$+$phonon many-particle states.  However, the wide absorption lines cover the virtual MP bands.

\subsection{PTCDA}
\label{subsec:ptcda}
The experimental linear absorption lines for the PTCDA crystal are broader than those of the MePTCDI \cite{hoffmann00}.  Thus, the structure of the PTCDA spectra does not exhibit many details generated by the FE--CTEs mixing.

Our calculations show the appearance of the spectral doublets in the excitonic and vibronic spectra associated with the FE--CTEs splitting (red curve 1 in Figure~10).  However, the probable value of the excitonic damping parameter is $\delta = 500$ cm$^{-1}$ evaluated using the width of the lowest experimental absorption maximum \cite{hoffmann00}.  The blue curve 2 in Figure~10 calculated with $\delta = 500$ cm$^{-1}$ exhibits four Lorentz-type maxima as it is observed in the experimental absorption spectra of the PTCDA crystal \cite{hoffmann00}.  The fifth maximum near $17\,000$ cm$^{-1}$ (see Figure~10) manifests itself as a weak peculiarity only of the absorption curve with $\delta = 500$ cm$^{-1}$.
\begin{figure}[!ht]
\centering\includegraphics[height=.25\textheight]{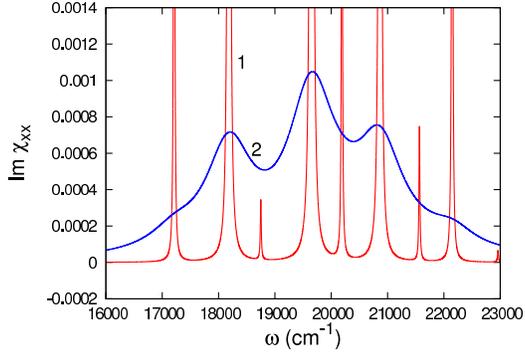}
  \caption{Linear absorption of the PTCDA crystal calculated using the ``exc'' program with $\xi_{\rm F} = 0.82$, $\xi = \xi_{\rm F}/2$, $\delta = 10$ cm$^{-1}$ (red curve 1), and $\delta = 500$ cm$^{-1}$ (blue curve 2).}
  \label{fig:fg10}
\end{figure}
\begin{figure}[!ht]
\centering\includegraphics[height=.25\textheight]{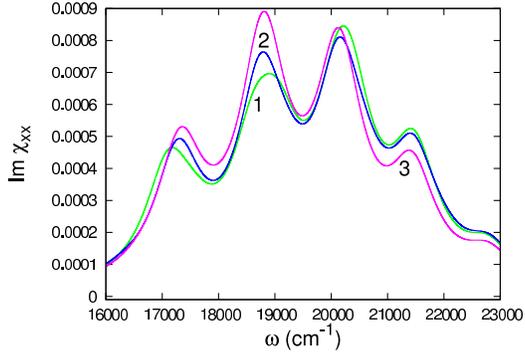}
  \caption{Linear absorption of the PTCDA crystal (``exc'' program) with $\xi_{\rm F} = 1.1$, $\delta = 500$ cm$^{-1}$,
  $\xi = \xi_{\rm F}/2$ (green curve 1) and $\xi = \xi_{\rm F}$ (blue curve 2).  The purple curve 3 corresponds to pure FE spectra ($\varepsilon_{\rm e}^{\prime \prime} = \varepsilon_{\rm h}^{\prime \prime} = 0$).}
  \label{fig:fg11}
\end{figure}
\begin{figure}[!ht]
\centering\includegraphics[height=.25\textheight]{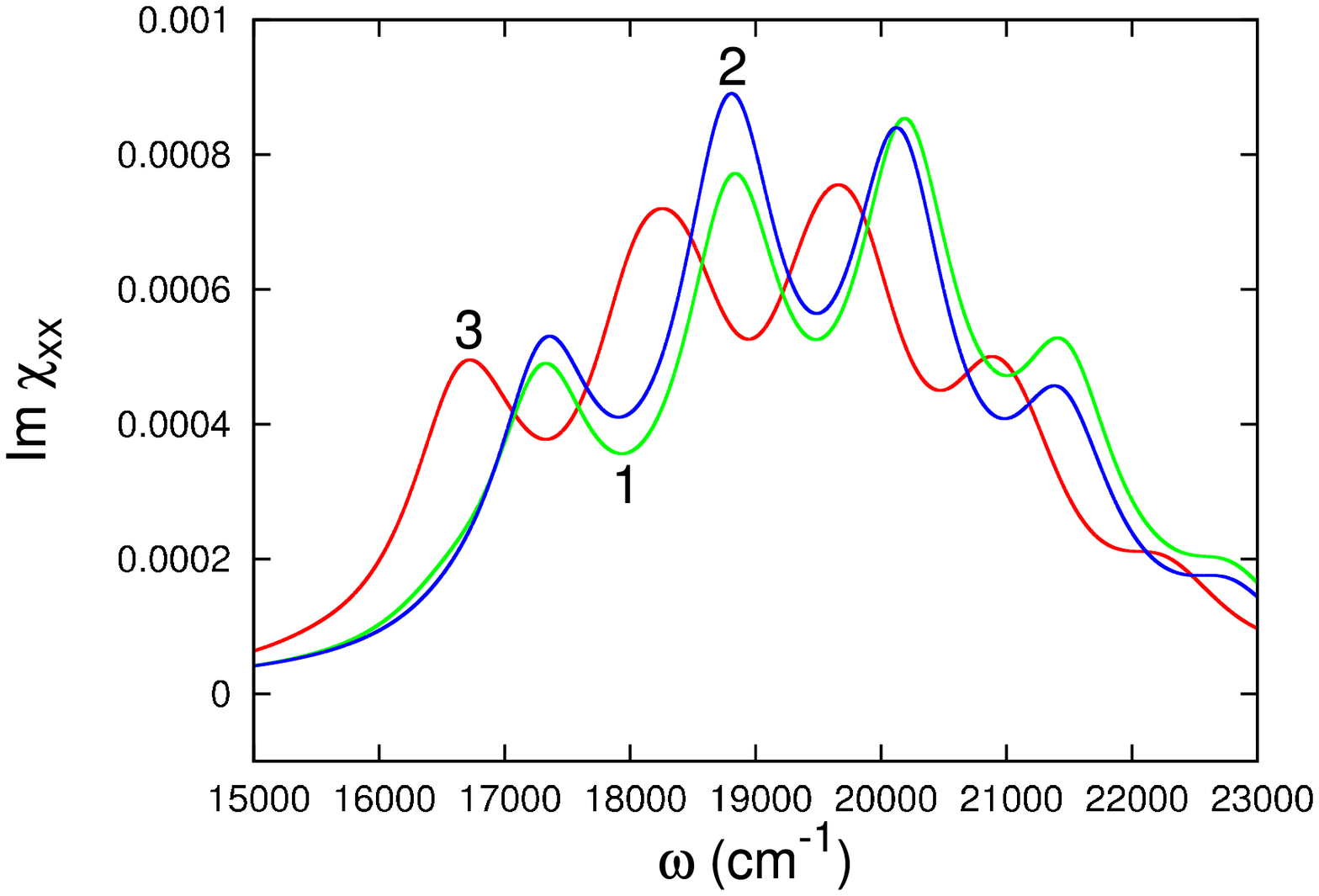}
  \caption{Linear absorption of the PTCDA crystal (``exc'' program) with $\xi_{\rm F} = 1.1$, $\xi = \xi_{\rm F}/2$, and $\delta = 500$ cm$^{-1}$.  The green curve 1 corresponds to $E_{\rm F} = 17\,992$ cm$^{-1}$, $E_{\rm c} = 17\,346$ cm$^{-1}$, the blue curve 2 to pure FE spectra ($\varepsilon_{\rm e}^{\prime \prime} = \varepsilon_{\rm h}^{\prime \prime} = 0$), and the red curve 3 to $E_{\rm F} = 17\,346$ cm$^{-1}$, $E_{\rm c} = 17\,992$ cm$^{-1}$.}
  \label{fig:fg12}
\end{figure}

Moreover, further calculations implemented with $\xi_{\rm F} = 1.1$ (which value seems to be the most suitable to our simulations) demonstrate insensitivity of the absorption spectra on the value of $\xi$ (see the very close green curve 1 and blue curve 2 in Figure~11).  We also calculated the absorption curve neglecting the FE--CTEs mixing (purple curve 3).  The result consists of some re-distributions of the absorption intensity among four maxima but the absorption lineshape is practically non-changed in the cases with and without FE--CTEs mixing.

The same effect -- the lineshape being not strongly affected by a hypothetical FE--CTEs mixing -- can be seen in Figure~12 (green curve 1 and blue curve 2 there).  The mutual replacement of the two excitonic levels (compare the red and green curves) shifts the position of the absorption maxima, however, their lineshapes are very similar.
\begin{figure}[ht]
  \begin{minipage}[b]{0.5\linewidth}
  \centering
    \includegraphics[width=5.3cm]{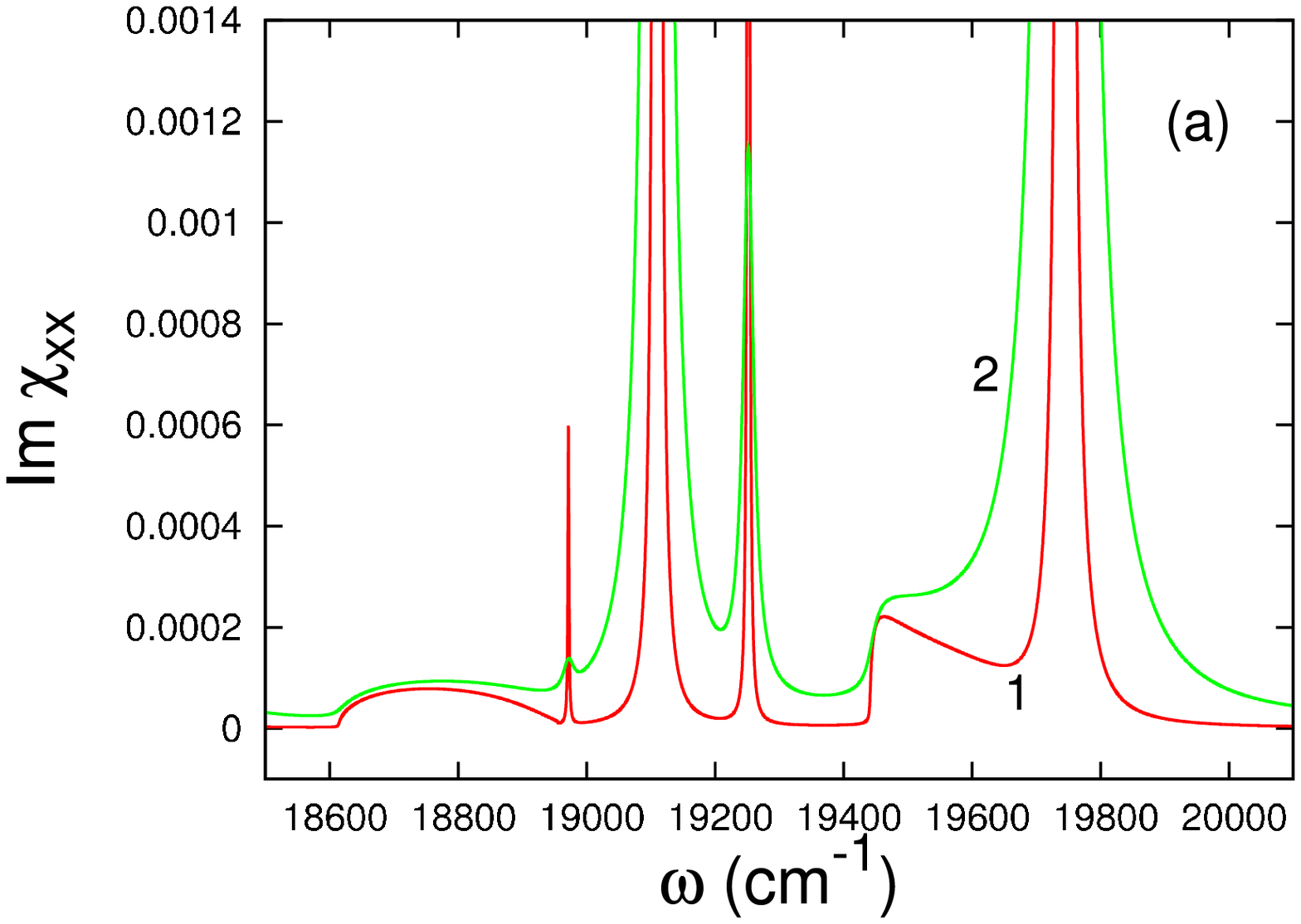}
  \end{minipage}
  \hspace{-0.2cm}
  \begin{minipage}[b]{0.5\linewidth}
  \centering
    \includegraphics[width=5.6cm]{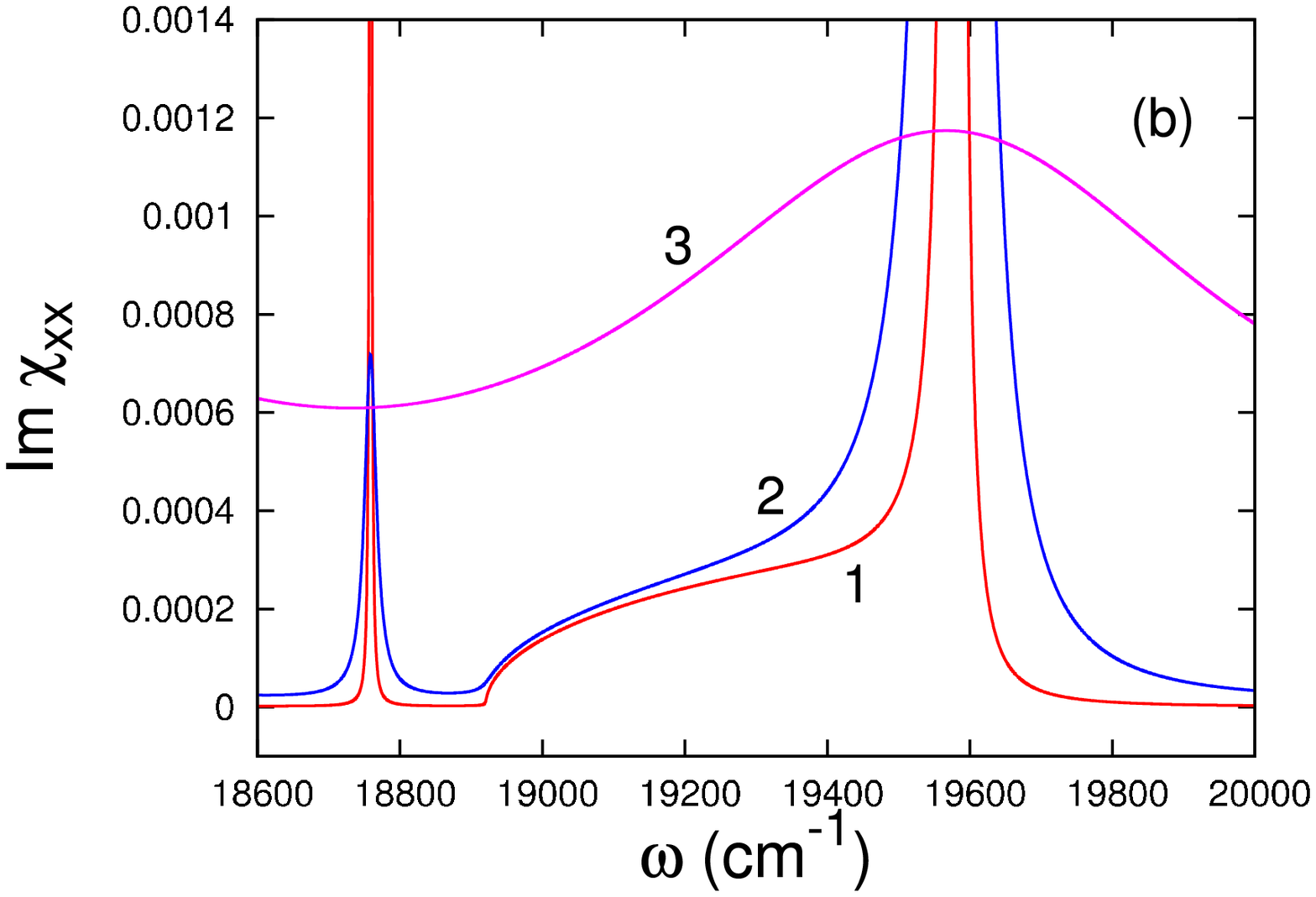}
  \end{minipage}
  \caption{One-phonon vibronic spectra of the PTCDA crystal (``1p'' program) with (a) $\xi_{\rm F} = 0.82$, $\xi = \xi_{\rm F}/2$, $\delta = 1$ cm$^{-1}$ (red curve 1), $\delta = 10$ cm$^{-1}$ (green curve 2) and (b) $\xi_{\rm F} = 0.82$, $\xi = \xi_{\rm F}/2$, and $\varepsilon_{\rm e}^{\prime \prime} = \varepsilon_{\rm h}^{\prime \prime} = 0$ (vibronic spectra of FE).  The red curve 1 corresponds to $\delta = 1$ cm$^{-1}$, the blue curve 2 to $\delta = 10$ cm$^{-1}$, and the purple curve 3 to $\delta = 500$ cm$^{-1}$.}
  \label{fig:fg13}
\end{figure}
\begin{figure}[!ht]
\centering\includegraphics[height=.25\textheight]{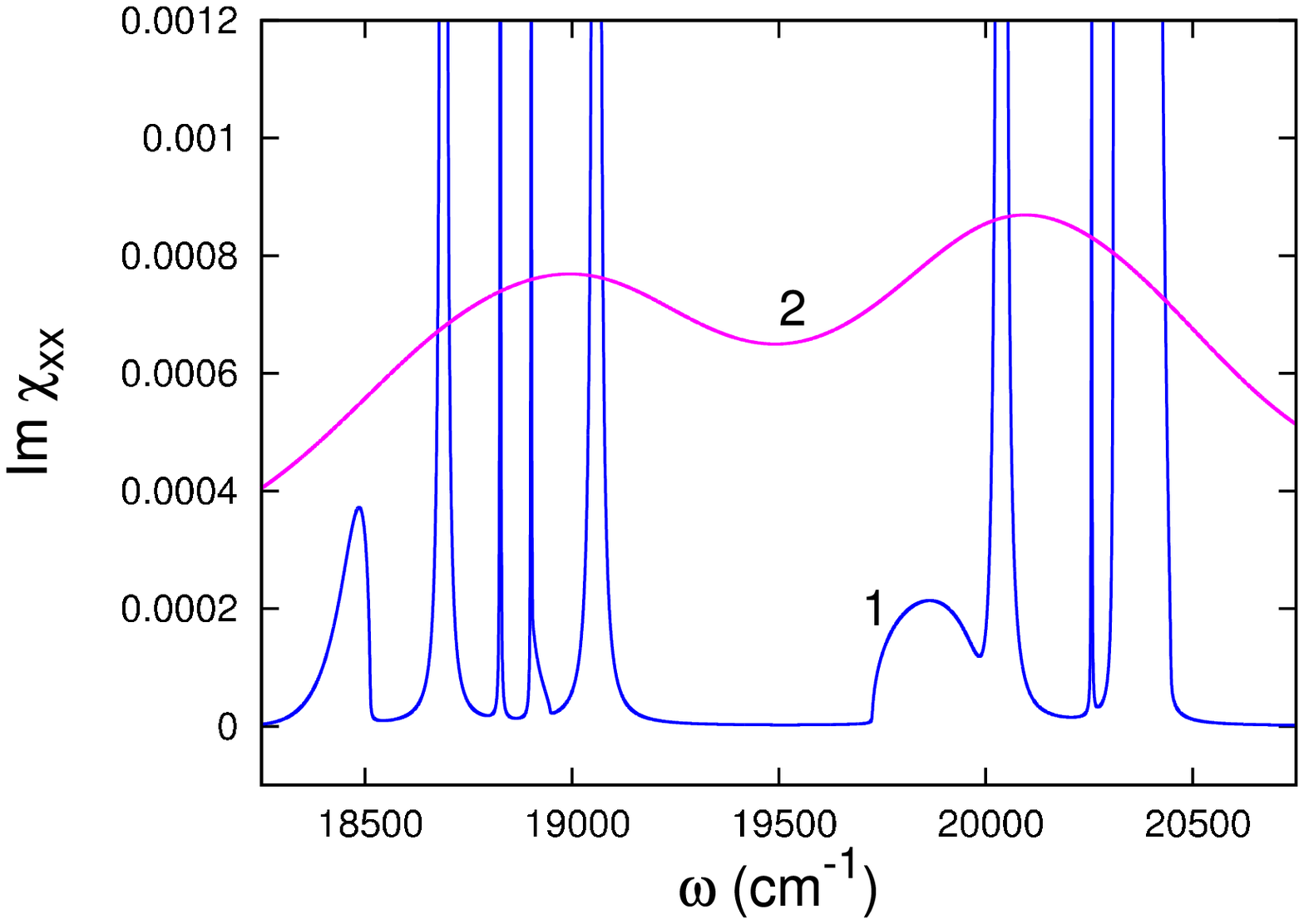}
  \caption{One-phonon and two-phonon vibronic spectra of the PTCDA crystal (``1p'' program) with $\xi_{\rm F} = 1.1$, $\xi = \xi_{\rm F}/2$.  The blue curve 1 corresponds to $\delta = 1$ cm$^{-1}$ and the purple curve 2 to $\delta = 500$ cm$^{-1}$.}
  \label{fig:fg14}
\end{figure}

The absorption curves in Figures~13 and 14 have been calculated using the ``1p'' program and they show the positions and the structure of one-phonon and two-phonon vibronic spectra.  Surely, the fine structure can be observed in the hypothetical case of small values of $\delta$.  Then the structure of the MP bands depends on the FE--CTEs mixing.  Figure~13(a) containing two absorption regions, $18\,600$--$19\,200$ cm$^{-1}$ and $19\,400$--$19\,900$ cm$^{-1}$, each one with a weak MP band and a Lorentzian.  This picture is rather different in the case of the absence of FE--CTEs mixing (Figure~13(b)).  Then we might observe a CTEs vibronic maximum near $18\,800$ cm$^{-1}$ and the vibronic replica of FE with MP band ($18\,900$--$19\,400$ cm$^{-1}$), as well as a Lorentz-type maximum at $19\,600$ cm$^{-1}$.  In our model Hamiltonian (\ref{eq:cteshamilt}) the dispersion of pure CTEs has been neglected and thus at $\varepsilon_{\rm e}^{\prime \prime} = \varepsilon_{\rm h}^{\prime \prime} = 0$ the vibronic replica of CTEs consists of one-particle maximum only.  In the same time, FEs possess dispersion expressed through the transfer terms with $L$ in expression (\ref{eq:fehamilt}) (see also \cite{hoffmann00,schmidt02}) and in the vibronic absorption spectra of FEs both one-particle maxima and MP band could appear.  As in other cases, a strong damping with $\delta = 500$ cm$^{-1}$ masks the whole structure.

The absorption curves in Figure~14 correspond to a higher value of $\xi_{\rm F} = 1.1$.  The one-phonon vibronic spectra ($18\,500$--$19\,000$ cm$^{-1}$) and two-phonon vibronic spectra ($19\,700$--$20\,500$ cm$^{-1}$) contain MP bands ($18\,300$--$18\,500$ cm$^{-1}$ and $19\,700$--$20\,000$ cm$^{-1}$) and several other maxima.  Their splitting is caused by the relatively big value of the constant $\xi_{\rm F}$ of the linear FE--phonon coupling.  In the realistic case of strong damping, $\delta = 500$ cm$^{-1}$, the one-phonon and two-phonon vibronic spectra would contain two broad and asymmetric maxima only.

The main conclusion of our simulations of the linear absorption in the PTCDA crystal is that the hypothesis of a FE--CTEs mixing is not crucial in describing the observed spectra.  The wide absorption lines in the excitonic and vibronic spectra cover the effects of the mixing and the hypothesis of Frenkel excitons' spectra and the vibronic of FE could be sufficient (see Ref.~\cite{vragovic03}).

\section{Conclusion}
\label{sec:concl}
Our model for the linear absorption spectra of the one-component charge-transfer molecular crystals includes the following parameters:
\begin{itemize}
\itemsep 0pt \parskip 0pt
\item excitonic levels $E_{\rm F}$ and $E_{\rm c}$, as well as the vibrational frequency $\omega_0$ of the intramolecular mode,
\item parameters $\varepsilon_{\rm e}^{\prime \prime}$ and $\varepsilon_{\rm h}^{\prime \prime}$ of the FE--CTEs mixing,
\item constants $\xi_{\rm F}$ and $\xi$ of the linear exciton--phonon coupling,
\item the width $\delta$ of the excitonic linewidth,
\item angles $2\varphi$, $\gamma$ and the ratio $a$ of the CTEs and FE transition dipole moments.
\end{itemize}

Practically all parameters have been introduced and fitted for MePTCDI and PTCDA crystals in previous papers \cite{henessy99,hoffmann00,hoffmann02,schmidt02}, based primarily on the matrix diagonalization method.  In our study we apply the complex vibronic approach based on the Green functions formalism in calculating the linear optical susceptibility and its imaginary part which is a factor in the absorption coefficient in the excitonic and one-phonon vibronic spectra.  Higher vibronics -- with two, three phonons -- also have been demonstrated in our calculations.  Calculated vibronic spectra consist of one-particle Lorentzian maxima and many-particle (MP) bands which correspond to unbound propagation of the excitons and phonons.  Our simulations expose the positions of the MP continua  but the wide excitonic and vibronic absorption lines in both crystals cover the fine structure of the vibronic spectra.
The main goal of our model is to simulate the lineshape in the absorption region of $15\,000$--$23\,000$ cm$^{-1}$ of the aforementioned crystals, and to find out better fitted values of the excitonic linewidth, exciton--phonon coupling parameters and so on.

We stress again the main results which concern the two crystals:
\begin{enumerate}
\itemsep 0pt \parskip 0pt
\item The excitonic linewidth of the MePTCDI crystal, according to our simulations, is approximately $\delta \approx 300$ cm$^{-1}$.  The FE-phonon linear coupling coefficient $\xi_{\rm F}$ has been estimated correctly in the previous papers \cite{henessy99,hoffmann02} as $\xi_{\rm F} \approx 0.88$ and $\xi = \xi_{\rm F}/2$.  The absorption spectra depend strongly on the mutual position of the two excitonic levels $E_{\rm F}$ and $E_{\rm c}$.  A supposition for the FE--CTEs mixing \emph{is necessary\/} for an adequate interpretation of the excitonic and vibronic spectra of the MePTCDI crystal.
\item The excitonic linewidth of the PTCDA crystal can be evaluated as $\delta \approx 500$ cm$^{-1}$.  That is why the effect of the FE--CTEs mixing, being covered by the wide absorption maxima, are more weakly expressed than in the MePTCDI crystal.  However, a very probable conclusion from our calculations may be the stronger linear exciton--phonon coupling ($\xi_{\rm F} \approx 1$ or $1.1$ instead of $0.82$).
\end{enumerate}

Our model can be applied in the interpretation of other one-component molecular stacks (crystals).  It can be more effective in the systems with narrower excitonic absorption lines where both types of vibronic states -- one-particle states and MP continua -- calculated by using the Green functions formalism would be seen in the linear absorption.

\appendix
\section*{Appendix}
\label{sec:app}
\setcounter{equation}{1}
\renewcommand{\theequation}{A.\arabic{equation}}
In this Appendix, we treat two groups of problems which concern the mixing of Frenkel and charge-transfer excitons in an one-component stack of the $C_i$ point group of symmetry.

(A) The first group is rather formal.  In the $C_i$ point group, at the center of Brillouin zone, $k = 0$, electronic excitations of the stack possess the following characteristics:
\begin{enumerate}
\renewcommand{\labelenumi}{\arabic{enumi}.}
\itemsep 0pt \parskip 0pt
\item The symmetry of inversion center makes the parity of the excitations odd or even and produces \emph{ungerade\/} and \emph{gerade\/} excitons.  In another part of the Brillouin zone these two types of excitons can mix producing some interesting effects (see, for example, Refs.~\cite{slawik97} and \cite{lalov08}).
\item The ungerade excitons only are dipole active and possess non-vanishing transition dipole moment.  For CTEs the ungerade wave function is the combination \cite{haarer75,sebastian81,sebastian83}:
    \begin{equation}
    \label{eq:ungeradewf}
        \frac{1}{\sqrt{2}}\left( U^{+}_{2,k=0} - U^{+}_{1,k=0} \right)|0\rangle = \frac{1}{\sqrt{2N}} \sum_n \left( U^{+}_{2,n} - U^{+}_{1,n} \right)|0\rangle.
    \end{equation}
\item The mixing of two excitons of different types of FE and CTEs is allowed only in the case of the same parity of the coupling excitons.

    Thus, the ungerade FEs as well as the ungerade CTEs only can mix and can be active in the linear absorption.  In the same time, the gerade FE and the gerade combination of CTEs
    \begin{equation}
    \label{eq:geradewf}
        \frac{1}{\sqrt{2}}\left( U^{+}_{2,k=0} + U^{+}_{1,k=0} \right)|0\rangle = \frac{1}{\sqrt{2N}} \sum_n \left( U^{+}_{2,n} + U^{+}_{1,n} \right)|0\rangle.
    \end{equation}
    can mix but they will be non-active in the linear absorption.
\end{enumerate}

(B) The second group of problems concerns the mixing constants $\varepsilon_{\rm e1}$, $\varepsilon_{\rm e2}$, $\varepsilon_{\rm h1}$, and $\varepsilon_{\rm h2}$ in a molecular stack of the $C_i$ point group (see Eqs.~(\ref{eq:fcoperator}) and (\ref{eq:h1new}) in the text).

Let us consider three neighbour molecules A, B, and D located at the positions $n-1$, $n$, and $n+1$ of the stack: $A_{n-1}$, $B_{n}$, $D_{n+1}$.  We denote by $B_n^{\ast}$ a neutral excited molecule (with FE on it) and by $A^{\pm}$, $B^{\pm}$, $D^{\pm}$ -- the wave functions of the corresponding positive and negative ions.  The matrix elements of the electron transfer can be represented as follows:
\begin{eqnarray}
\label{eq:epse1}
    \langle A_{n-1} B_n^{\ast} D_{n+1} |\hat{H}| A_{n-1} B_n^{+} D^{-}_{n+1} \rangle \nonumber \\
    {}\equiv \langle A_{n-1} B_n^{\ast} D_{n+1} |\hat{H}| A_{n-1}C_{n,1} \rangle = \varepsilon_{\rm e1},
\end{eqnarray}
\begin{eqnarray}
\label{eq:epse2}
    \langle A_{n-1} B_n^{\ast} D_{n+1} |\hat{H}| A^{-}_{n-1} B_n^{+} D_{n+1} \rangle \nonumber \\
    {}\equiv \langle A_{n-1} B_n^{\ast} D_{n+1} |\hat{H}| C_{n,2}D_{n+1} \rangle = \varepsilon_{\rm e2},
\end{eqnarray}
where $\hat{H}$ is the Hamiltonian, $C_{n,\sigma}$ are the wave functions of the CTEs, $\varepsilon_{\rm e1}$ and $\varepsilon_{\rm e2}$ are the mixing constants (similar expressions hold for $\varepsilon_{\rm h1}$ and $\varepsilon_{\rm h2}$).

The gerade FE ($B_n^{\ast}$ is an even function relatively to the symmetry of inversion) mixes with the combination $(1/\sqrt{2})\left( C^{+}_{n,1} + C^{+}_{n,2} \right)|0\rangle$ and the mixing constant is
\begin{equation}
\label{eq:mixconst1}
    \frac{1}{\sqrt{2}} \left( \varepsilon_{\rm e1} + \varepsilon_{\rm e2} \right).
\end{equation}
In the case of ungerade FE the mixing constant with the ungerade combination $(1/\sqrt{2})\left( C^{+}_{n,2} - C^{+}_{n,1} \right)|0\rangle$ is:
\begin{equation}
\label{eq:mixconst2}
    \frac{1}{\sqrt{2}} \left( \varepsilon_{\rm e2} - \varepsilon_{\rm e1} \right).
\end{equation}

The manifestation of the FE--CTEs mixing depends on the conditions at which the following equalities hold:
\begin{equation}
\label{eq:equalities}
    \varepsilon_{\rm e1} = \varepsilon_{\rm e2} \qquad \mbox{and} \qquad  \varepsilon_{\rm h1} = \varepsilon_{\rm h2}.
\end{equation}
The disposition of the molecules in the stack allows two possible configurations in which relations~(\ref{eq:equalities}) are fulfilled:
\begin{enumerate}
\renewcommand{\labelenumi}{\arabic{enumi})}
\itemsep 0pt \parskip 0pt
\item The planes of the molecules are perpendicular to the stack's axis.  In that case the configurations $B^{+}_n D^{-}_{n+1}$ and $A^{-}_{n-1}B^{+}_n$ are mirror-symmet\-rical with respect to the positive ion $B^{+}_n$ and hence the quantities (\ref{eq:epse1}) and (\ref{eq:epse2}) are equal, independently on the symmetry of the molecular orbitals in the positive ion and, on the other hand, in the negative ion.  Obviously the perpendicular position of the molecules in the stack is not the case of the treated crystals \cite{vragovic03}.

\item In the case of inclination of the molecules' planes relatively to the stack's axis the full symmetry of the positive ion $B^{+}_n$, governed by the symmetry of the molecular orbitals on it, must coincide with the full symmetry of the negative ions $A^{-}_{n-1}$ and $D^{-}_{n+1}$ (governed by their molecular orbitals).  Especially the maxima of the charge distribution in the positive and negative ions must be located at the same points of the entity.  In the opposite case, because of the inclination, the conditions of the electron/hole transfer from the neutral excited molecule $n$ on the its two neighbours $n-1$ and $n+1$ are non-equivalent.
\end{enumerate}

Since situation 2) seems to be less probable than the non-coinciding symmetry of the positive and negative ions, the inequalities $\varepsilon_{\rm e1} \neq \varepsilon_{\rm e2}$ and $\varepsilon_{\rm h1} \neq \varepsilon_{\rm h2}$ hold.  This is notably the case of mixing FEs and CTEs which are active in the linear absorption, too.  In that case, formula (\ref{eq:fc}) describes the FE--CTEs mixing of the dipole active excitons.

In the same time, the following operator describes the FE--CTEs mixing of gerade excitons (in all the configurations):
\begin{eqnarray}
\label{eq:Hfc}
    \hat{H}_{\rm FC} = \sum_k \left[ \left( \varepsilon_{\rm e}^{\prime} + \varepsilon_{\rm h}^{\prime}\cos k \right) V_k^{+} \left( U_{k,1} + U_{k,2} \right) \right. \nonumber \\
    \left.
    {}+ \mathrm{i}\varepsilon_{\rm h}^{\prime}\sin k V_k^{+} \left( U_{k,2} - U_{k,1} \right) + \mbox{h.c.} \right]
\end{eqnarray}
(see formula (\ref{eq:epsprime}) for the expressions of $\varepsilon_{\rm e}^{\prime}$ and $\varepsilon_{\rm h}^{\prime}$).

The spectra of the gerade mixing FE--CTEs are similar to the spectra calculated in this paper, but they will not manifest themselves in the linear absorption.

If configurations 1) or 2) are realized, then relations (\ref{eq:equalities}) are fulfilled and operator (\ref{eq:Hfc}) describes the FE--CTEs mixing among gerade excitons only.  In this case, the spectra of the linear absorption consist of two independent (non-mixing) spectra, notably: the linear absorption spectra of FEs and their vibronics, and the spectra of the CTEs and their vibronics.  The simulation of non-mixing FEs and CTEs can be performed by using the formulas in the present paper, in which we have to put $\varepsilon_{\rm e}^{\prime \prime} = \varepsilon_{\rm h}^{\prime \prime} = 0$.


\begin{thebibliography}{26}
\bibitem{sebastian81}       
L.~Sebastian,  G.~Weisser, G.~Peter, and H.~B\"{a}ssler (1981) \emph{Chem.\ Phys.}\ \textbf{61} 125.

\bibitem{sebastian83}       
L.~Sebastian,  G.~Weisser, G.~Peter, and H.~B\"{a}ssler (1983) \emph{Chem.\ Phys.}\ \textbf{75} 103.

\bibitem{siebrand83}        
W.~Siebrand and M.~Z.~Zgierski (1983) in \emph{Organic Molecular Aggregates}, Springer Series in Solid-State Sciences, Vol.~18, eds.~P.~Reineker, H.~Haken, and M.C.~Wolf, Springer, Berlin, p.~136.

\bibitem{petelenz96}        
P.~Petelenz, M.~Slawik, K.~Yokoi, and M.~Zgierski (1996) \emph{J.\ Chem.\ Phys.}\ \textbf{105} 4427.

\bibitem{henessy99}         
M.~H.~Henessy, Z.~G.~Soos, R.~A.~Pascal Jr., and A.~Girlando (1999) \emph{Chem.\ Phys.}\ \textbf{245} 199.

\bibitem{hoffmann00}        
M.~Hoffmann, K.~Schmidt, T.~Fritz, V.~M.~Agranovich, and K.~Leo (2000) \emph{Chem.\ Phys.}\ \textbf{258} 73.

\bibitem{hoffmann02}        
M.~Hoffmann and Z.~G.~Soos (2002) \emph{Phys.\ Rev.}\ B \textbf{66} 024305.

\bibitem{jeglinski92}       
S.~Jeglinski, Z.~V.~Vardeny, D.~Moses, V.~I.~Srdanov, and F.~Wudl (1992) \emph{Synth.\ Met.}\ \textbf{49-50} 557.

\bibitem{pac98}             
B.~Pac, P.~Petelenz, A.~Eilms, and R.~W.~Munn (1998) \emph{J.\ Chem.\ Phys.}\ \textbf{109} 7932.

\bibitem{schmidt02}         
K.~Schmidt, K.~Leo, and V.~M.~Agranovich (2002) in \emph{Organic Nanonstructures: Science and Applications}, eds.~V.~M.~Agranovich and G.~C.~La Rocca, IOS Press, Amsterdam, p.~521.

\bibitem{lalov06a}          
I.~J.~Lalov and I.~Zhelyazkov (2006) \emph{Chem.\ Phys.}\ \textbf{321} 223.

\bibitem{lalov07a}          
I.~J.~Lalov, C.~Supritz, and P.~Reineker (2007) \emph{Chem.\ Phys.}\ \textbf{332} 108.

\bibitem{lalov06b}          
I.~J.~Lalov and I.~Zhelyazkov (2006) \emph{Chem.\ Phys.\ Res. J.}\ \textbf{1} 75.

\bibitem{lalov06c}          
I.~J.~Lalov and I.~Zhelyazkov (2006) \emph{Phys.\ Rev.}\ B \textbf{74} 035403.

\bibitem{lalov07b}          
I.~J.~Lalov and I.~Zhelyazkov (2007) \emph{Phys.\ Rev.}\ B \textbf{75} 245435.

\bibitem{lalov08}           
I.~J.~Lalov, C.~Warns, and P.~Reineker (2008) \emph{New J.\ Phys.}\ \textbf{10} 085006.

\bibitem{haarer75}          
D.~Haarer, M.~R.~Philpott, and H.~Morawitz (1975) \emph{J.\ Chem.\ Phys.}\ \textbf{63} 5238.

\bibitem{brillante80}       
A.~Brillante and M.~R.~Philpott (1980) \emph{J.\ Chem.\ Phys.}\ \textbf{72} 4019.

\bibitem{weiser04}          
G.~Weiser (2004) \emph{J.\ Lumin.}\ \textbf{110} 189.

\bibitem{lalov05}           
I.~J.~Lalov, C.~Supritz, and P.~Reineker (2005) \emph{Chem.\ Phys.}\ \textbf{309} 189.

\bibitem{davydov71}         
A.~S.~Davydov (1971) \emph{Theory of Molecular Excitons}, Plenum, New York.

\bibitem{agranovich83}      
V.~M.~Agranovich (1983) in \emph{Spectroscopy and Exciton Dynamics of Condensed Molecular Systems}, eds.~V.~M.~Agranovich and R.~M.~Hochstrasser, North Holland, Amsterdam, p.~83.

\bibitem{note}              
In the crystals under consideration vectors $\mathbf{p}_{\rm CT}^{A}$ and $\mathbf{p}_{\rm CT}^{ B}$ are situated very close to the same plane ($XY$), see \cite{hoffmann00}.

\bibitem{haedicke86}        
E.~H\"{a}dicke and F.~Graser (1986) \emph{Acta Cryst.}\ C \textbf{42} 189.

\bibitem{vragovic03}        
I.~Vragovi\'c and R.~Scholz (2003) \emph{Phys.\ Rev.}\ B \textbf{68} 155202.

\bibitem{slawik97}          
M.~Slawik and P.~Petelenz (1997) \emph{J.\ Chem.\ Phys.}\ \textbf{107} 7114.
\end{thebibliography}
\end{document}